\documentclass{jfm}

\usepackage{graphicx}
\usepackage{natbib}
\usepackage{epstopdf}
\usepackage{amsmath}
\usepackage{esdiff}
\usepackage{subfig}
\usepackage{float}

\ifCUPmtlplainloaded \else
  \checkfont{eurm10}
  \iffontfound
    \IfFileExists{upmath.sty}
      {\typeout{^^JFound AMS Euler Roman fonts on the system,
                   using the 'upmath' package.^^J}%
       \usepackage{upmath}}
      {\typeout{^^JFound AMS Euler Roman fonts on the system, but you
                   dont seem to have the}%
       \typeout{'upmath' package installed. JFM.cls can take advantage
                 of these fonts,^^Jif you use 'upmath' package.^^J}%
      }
  \else
  \fi
\fi

\ifCUPmtlplainloaded \else
  \checkfont{msam10}
  \iffontfound
    \IfFileExists{amssymb.sty}
      {\typeout{^^JFound AMS Symbol fonts on the system, using the
                'amssymb' package.^^J}%
       \usepackage{amssymb}%
         \let\leq=\leqslant
         
      }{}
  \fi
\fi

\ifCUPmtlplainloaded \else
  \IfFileExists{amsbsy.sty}
    {\typeout{^^JFound the 'amsbsy' package on the system, using it.^^J}%
     \usepackage{amsbsy}}
    {}
\fi

\newsavebox{\astrutbox}
\sbox{\astrutbox}{\rule[-5pt]{0pt}{20pt}}

\title{Interaction Forces Between Microfluidic Droplets in a Hele-Shaw Cell}
\shorttitle{Interaction Between Droplets in a Hele-Shaw Cell}
\author[I. Sarig, Y. Starosvetsky, A.D. Gat]{I. Sarig, Y. Starosvetsky, A.D. Gat}
\affiliation{Faculty of Mechanical Engineering, Technion - Israel Institute of Technology, Haifa 3200003, Israel}

\pubyear{2015} \volume{999} \pagerange{999--9999}
\date{2015}

\begin{document}

\maketitle


\begin{abstract}
Various microfluidic systems, such as chemical and biological lab-on-a-chip devices, involve motion of multiple droplets within an immersing fluid in shallow micro-channels. Modeling the dynamics of such systems requires calculation of the forces of interaction between the moving droplets. These forces are commonly approximated by superposition of dipoles solutions, which requires an assumption of sufficiently large distance between the droplets. In this work we obtain exact solutions (in the Hele-Shaw limit) for two moving droplets, and a droplet within a droplet, located within a moving immersing fluid, without limitation on the distance between the droplets. This is achieved by solution of the pressure field in a bi-polar coordinate system and calculation of the force in a Cartesian coordinate system. Our results are compared with numerical computations, experimental data and the existing dipole-based models. We utilize the results to calculate the dynamics of a droplet within a droplet, and of two close droplets, located within an immersing fluid with oscillating speed. The obtained results may be used to study the dynamics of dense droplet lattices, common to many microfluidic systems.
\end{abstract}

\section{Introduction}
 The motion of multiple droplets contained within a moving immersing fluid in a Hele-Shaw geometry is relevant to various micro-fluidic systems, such as chemical and biological lab-on-a-chip devices \citep{stone2004engineering,squires2005microfluidics,zhao2013multiphase}. Generation of such configurations is commonly done by controlled injection, via T-junction, of one liquid into a second immiscible
liquid \citep[][among many others]{Garstecki,christopher2008experimental,desreumaux2013hydrodynamic}. 

Since inertial effects and gravity are commonly negligible due to the small size associated with microfluidic configurations, the flow-field is governed by surface related effects such as capillarity and viscosity. Modeling the dynamics of such systems requires calculation of the forces of interaction between multiple droplets moving relatively to each-other and the surrounding fluid. These forces are commonly approximated by superposition of dipole solutions \citep[][]{beatus2006phonons,uspal2012collective,shani2013long,fleury2014mode}, which requires an assumption of sufficiently large distance between the droplets. Given the lack of models of interaction forces between closely spaced droplets, the dipole based models are used also for cases which violate the assumption of long distances between the droplets \citep[e.g.][]{Beatus2012103,liu2012waves,shen2014dynamics}. In addition, some systems involve droplet within a droplet configurations \citep{hindmarsh2005pfg,he2005selective,hanson2008nanoscale}, which have not been previously modelled in a Hele-Shaw geometry, to the best of our knowledge.

The aim of this work is to obtain exact solutions for interaction forces between two relatively moving droplets, and a droplet within a droplet, located within a moving immersing fluid in a Hele-Shaw cell, without limitation on the distance between the droplets.

This paper is organized as follows. In \S2 we describe the microfluidic droplet configuration and relevant assumptions made in the model. In \S3 we obtain the pressure field created by two relatively moving, closely-spaced, droplets in the bi-polar coordinate system. In \S4 we calculate the interaction forces between two droplets in the Cartesian coordinate system. We compare our model to experimental results, the existing dipole model, as well as numerical computations. We then apply our results to examine the dynamics of two closely-spaced droplets, and a droplet within a droplet, subjected to the external oscillating velocity.  In \S5 we give concluding remarks.

\section{Problem Formulation}
\begin{figure}
\centering
\includegraphics[width=12.cm]{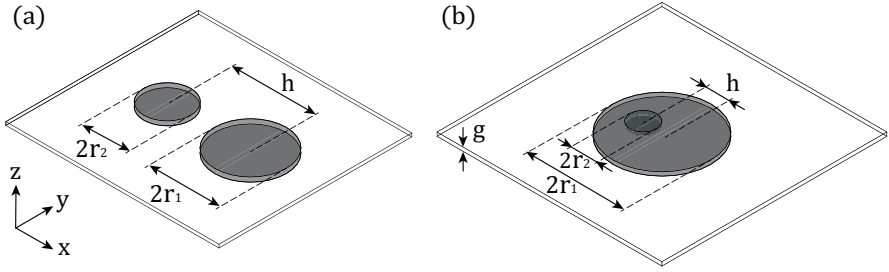}
\caption{Schematic description of the droplets configuration and coordinate system. The coordinate $x$ is defined as parallel to a line connecting the centers of the droplets. Panel $(a)$ presents two closely spaced droplets and panel $(b)$ presents a droplet within a larger droplet. $r_1,r_2$ are the radii of droplets $1$ and $2$, respectively. $h$ is the distance between the centers of the droplets, $g$ is the gap between the upper and lower plates.}
\label{fig:ModelDescribtion}
\end{figure}

We focus on the interaction forces between two closely spaced droplets  (see Fig. \ref{fig:ModelDescribtion}a), or a droplet within a droplet (see Fig. \ref{fig:ModelDescribtion}b), positioned between two parallel flat plates separated by a small gap. The droplets are immersed within a different immiscible liquid flowing uniformly far from the droplets.

The Cartesian coordinate system is denoted by $(x,y,z)$ and time is denoted by $t$. The coordinate $x$ is defined as parallel to a line connecting the centers of the droplets. The upper and lower plates are parallel to the $x-y$ plane. The gap between the parallel plates is denoted by $g$ and is assumed to be small compared to all other length scales of the problem. The droplets are assumed circular in the $x-y$ plane due to dominant surface tension effects. The centers of the droplets $1$ and $2$ are denoted by $(x_1,y_1)$ and $(x_2,y_2)$, respectively. The distance between the centers of the droplets is denoted by $h$. The radii of droplets $1$ and $2$ are denoted by $r_1$ and $r_2$, respectively. The $z$-averaged liquid velocity in the $x-y$ plane is denoted by $(u,v)$ and the uniform $z$-averaged velocity of the surrounding liquid far from the location of the droplets is denoted by $(u_\infty, v_\infty)$. The velocities of droplets $1$ and $2$ are $(u_1, v_1)$ and $(u_2, v_2)$, respectively.  The viscosity of the surrounding liquid is $\mu_s$ and the viscosity of droplets $1$ and $2$ is denoted by $\mu_1$ and $\mu_2$, respectively. We focus on configurations with negligible inertial effects.

Hereafter, normalized variables will be denoted by capital letter and characteristic values will be denoted by asterisk superscript. We define the normalized coordinates $(X,Y,Z)$ and time $T$,
\begin{subequations}
   \begin{equation} \label{eq01}
         X=\frac{x}{l^*},\quad Y=\frac{y}{l^*},\quad Z=\frac{z}{g},\quad T=\frac{t}{l^*/u^*}, 
   \end{equation}
normalized $Z$-averaged velocity in the $X-Y$ plane and normalized pressure $P$,
      \begin{equation} \label{eq011}
U=\frac{u}{u^*},\quad V=\frac{v}{u^*},\quad P=\frac{p}{12\mu_i u^*l^*/g^2},
   \end{equation}
    \end{subequations}
where $\mu_i$ is the viscosity of the liquid in the region of calculation  (see Fig. \ref{fig:BiPolarCoordinates}b,c).
The normalized droplets radii and distance between the droplets centers are $(R_1,R_2)=(r_1/l,r_2/l)$ and $H=h/l^*$, respectively. The normalized velocities of the droplets and the velocity far from the droplets are $(U_1,V_1)=(u_1/u^*,v_1/u^*)$, $(U_2,V_2)=(u_2/u^*,v_2/u^*)$ and $(U_\infty,V_\infty)=(u_\infty/u^*,v_\infty/u^*)$, respectively.

Using the assumptions of negligible inertial effects and small gap between the plates, we apply the Hele-Shaw approximation, and thus the governing equation is the Laplace equation for the liquid pressure $\nabla_{||}^2P=0$, where $\nabla_{||}$ is the two-dimensional nabla operator in the $X-Y$ plane, and $\partial P/\partial Z=0$. The relation between the $Z$-averaged normalized velocity, $(U,V)$, and the normalized pressure gradient is $(U,V)=-\nabla_{||}P$. The Laplace equation is supplemented by the no-penetration condition at the boundary of the droplets,
\begin{equation} \label{eq1}
(\nabla_{||}P+(U_i,V_i))\cdot\hat{n}=0,
\end{equation}
where $\hat{n}$ is a unit vector normal to the boundary of the droplet and pointing outward,  $i=1,2$, and uniform velocity far from the droplets, $(U,V)\rightarrow (U_\infty,V_\infty)$.


\section{Calculation of Pressure Distribution in a Bi-Polar Coordinate System}
\begin{figure}
\centering
\includegraphics[width=13cm]{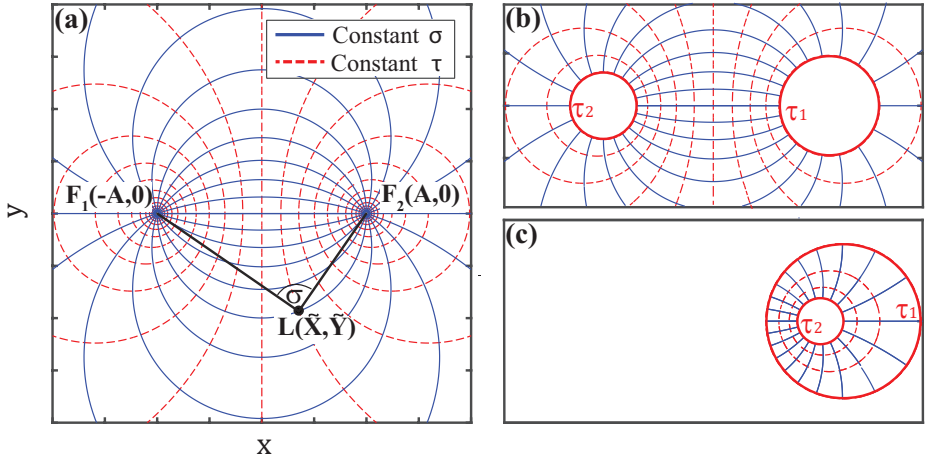}
\caption{Illustration of the bi-polar coordinate system. Panel $(a)$ presents isosurfaces of $\sigma$ (red dashed), isosurfaces of $ \tau$ (blue smooth) and two focal points  at $(-A,0),\ (A,0)$. $\sigma$ is the angle between the lines connecting point $L$ and the two focal points, and $\tau$ is the natural logarithm of the ratio of length of the lines connecting $L$ to the focal points. Panels $(b)$ and $(c)$ present two closely spaced droplets and a droplet within a droplet, respectively, in the $\sigma-\tau$ plane.
}
\label{fig:BiPolarCoordinates}
\end{figure}

We define auxiliary coordinates, denoted by tildes, as $(\tilde X,\tilde Y)$
\begin{equation}
 \tilde X=X+U_\infty T,\quad  \tilde Y=Y+V_\infty T,
\end{equation}
where the velocity of droplet $i$ is $(\tilde U_{i},\tilde V_{i})=(U_{i}-U_\infty,V_{i}-V_\infty)$ and $(\tilde U_{\infty},\tilde V_{\infty})=(0,0)$. We apply the  bi-polar coordinates transformation from the $\tilde X-\tilde Y$ plane to the $\sigma-\tau$ plane as described in Fig. \ref{fig:BiPolarCoordinates}a, where the curves of constant $\sigma$ (blue solid) and of constant $\tau$ (red dashed) are perpendicular and describe circles. The  $(\sigma,\tau)$ coordinates have two focal points, denoted by $F_1$ and $F_2$, located at $(A, 0)$ and $(-A, 0)$, respectively, on the $\tilde X$-axis of the Cartesian  coordinate system. The conformal transformation from the Cartesian coordinate system $(\tilde X,\tilde Y)$ to the bi-polar coordinates $(\sigma,\tau)$ is \citep{polyanin2001handbook}
\begin{equation} \label{eq3}
(\tilde X,\tilde Y) = A \left( \frac{\sinh \tau}{\cosh \tau - \cos \sigma}, \frac{\sin \sigma}{\cosh \tau - \cos \sigma}\right),
\end{equation}
where $\sigma$ of the point $L$ is the angle $\angle F_2LF_1$ and $\tau$ is the natural logarithm of the ratio of distances $|LF_1|$ and $|LF_2|$  respectively, $\tau=\ln{|LF_1|/|LF_2|}$.
Eq. (\ref{eq3}) can be presented as
\begin{equation} \label{eq: aeq4}
(\tilde X - A \cdot \coth \tau)^2 + \tilde Y^2 =  \frac{A^2}{\sinh^2 \tau},
\end{equation}
and thus the radius of droplet $i$ is given by $R_i = A/\sinh \tau$ and the droplet radius center is located at $A \coth \tau$. From (\ref{eq3}) $R_1$, $R_2$ and $H$ are related to  $\tau_1$, $\tau_2$ and $A$ (see Fig. \ref{fig:BiPolarCoordinates}a),
\begin{equation} \label{eq5}
\tau_1=\sinh ^{-1} \bigg(\frac{A}{R_1}\bigg),\quad \tau_2=C\sinh ^{-1} \bigg(\frac{A}{R_2}\bigg)
\end{equation}
and
\begin{equation} \label{eq6}
A= \frac{\sqrt{H^2-(R_1+R_2)^2)(H^2-(R_1-R_2)^2)}}{2H},
\end{equation}
where  $C=-1$ for two closely spaced droplets (Fig. \ref{fig:BiPolarCoordinates}b) and $C=1$ for a droplet within a droplet (Fig. \ref{fig:BiPolarCoordinates}c).

The governing equation in the $\sigma-\tau$ plane is
\begin{equation}\label{laplace_tau_sigma}
 \frac{\partial^2 \tilde{P}}{{\partial \sigma^2}} +\frac{\partial^2 \tilde{P}}{{\partial \tau^2}}=0,
\end{equation}
in the rectangular domain $\tau_1\leq\tau\leq\tau_2$ and $-\pi\leq\sigma\leq\pi$. Since $\sigma\rightarrow \pi$ and $\sigma\rightarrow -\pi$ both represent the line connecting the centers of the circles, from opposing directions, (\ref{laplace_tau_sigma}) is supplemented by the periodic boundary conditions at $\sigma=-\pi,\pi$
\begin{equation} \label{eq7}
\tilde P(\sigma = -\pi, \tau)= \tilde P(\sigma = \pi, \tau),\quad \frac{\partial \tilde P}{\partial \sigma} \bigg |_{\sigma = -\pi}= \frac {\partial \tilde P}{\partial \sigma}\bigg |_{\sigma =\pi},
\end{equation}
as well as no-penetration conditions on the boundary of the droplets at $\tau_i$ 
\begin{equation} \label{eq8}
\begin{split}
\frac{\partial P}{\partial \tau}\bigg|_{\tau=\tau_{i}}+\tilde U_{i} A \frac{1-\cosh {\tau_{i}}\cos{\sigma}}{(\cosh {\tau_{i}}- \cos{\sigma})^2}-\tilde V_{i}A \frac{\sin{\sigma}\sinh{\tau_{i}}}{(\cosh {\tau_{i}}- \cos{\sigma})^2}=0.
\end{split}
\end{equation}

Applying the separation of variables and solving the eigenvalue problem yields the general form of the solution
\begin{equation} \label{eq9}
\tilde{P}(\sigma, \tau)= \sum _{n=1} ^{\infty} {(\alpha_n \cos{n \sigma}+\beta_n \sin{n \sigma})(\gamma_n\cosh{n \tau}+\delta_n\sinh{n \tau})}.
\end{equation}
Utilizing (\ref{eq8}) on $\tau=\tau_i$ yields,
\begin{equation} \label{eq10}
\begin{split}
\alpha_n \delta_n=\frac{ -I_{1}  \sinh ⁡(n \tau_2 )+I_{2}  \sinh ⁡(n \tau_1 )}{n  \sinh⁡(n (\tau_{1}-\tau_{2} ))},\alpha_n\gamma_n=\frac{ I_{1}  \cosh ⁡(n \tau_2 )-I_{2}  \cosh (n \tau_1 )}{n  \sinh⁡(n (\tau_{1}-\tau_{2} ))}\\
\beta_n \gamma_n=\frac{ -J_{1}  \sinh ⁡(n \tau_2 )+J_{2}  \sinh (n \tau_1 )}{n  \sinh⁡(n (\tau_{1}-\tau_{2} ))},\beta_n \gamma_n=\frac{ J_{1}  \cosh ⁡(n \tau_2 )-J_{2}  \cosh (n \tau_1 )}{n  \sinh⁡(n (\tau_{1}-\tau_{2} ))}
\end{split}
\end{equation}
where
\begin{subequations} \label{eq11}
\begin{equation}
I_{i} = \frac {\tilde U_{i} A}{\pi} \int _{-\pi} ^{\pi} {\cos{n \sigma} \frac{\cosh{\tau _{i}}\cos{\sigma}-1}{(\cosh {\tau _{i}} - \cos {\sigma})^2} d\sigma}=2 \tilde{U}_i A n e^{-n |\tau _i|}
\end{equation}
\begin{equation}
J_{i} = \frac {\tilde V_{i} A}{\pi} \int _{-\pi} ^{\pi} {\sin{n \sigma} \frac{\sinh{\tau _{i}}\sin{\sigma}}{(\cosh {\tau _{i}} - \cos {\sigma})^2} d\sigma}= \begin{cases} 2 \tilde{V}_i A n e^{-n \tau _i} & \mbox{if } \tau_{i}>0 \\ -2 \tilde{V}_i A n e^{n \tau _i} &\mbox{if } \tau_{i}<0.
\end{cases}
\end{equation}
\end{subequations}
The pressure field in the $(X,Y)$ coordinates is thus obtained from (\ref{eq9})-(\ref{eq11}) together with (\ref{eq3}) and the transformation
\begin{equation} \label{eq:inverseP}
    P(X, Y)=\tilde P(\tilde X, \tilde Y)+ U_\infty X +V_\infty Y.
\end{equation}

\section{Interaction Forces and Dynamics of Two Closely Spaced Droplets and a Droplet within a Droplet}
In order to calculate the force acting on droplet $i$ in the Cartesian coordinates $(X,Y)$ we integrate 
\begin{equation} \label{eq:general forces}
 (F_{X,i},F_{Y,i})= - \oint _{droplet} {P( X, Y) \bigg(\frac{\partial  Y}{\partial S},\frac{\partial  X}{\partial S}\bigg) dS},
\end{equation}
where $S$ is a coordinate across the droplet boundary. In the $\sigma-\tau$ plane, $S$ corresponds the variation $\sigma$ for the fixed value of $\tau$. Thus the Jacobian is simply reads $\partial S/\partial \sigma$. The dependence of the Jacobian in $S$ eliminates its dependence in the reflection of the pressure in both $X,Y$ directions leading to the simplified expression 
\begin{equation} \label{eq:general_forces}
 (F_{X,i},F_{Y,i})=-\int _{-\pi} ^{\pi}{\tilde P(\sigma, \tau _{1})  \bigg(\frac{\partial  Y}{\partial \sigma},\frac{\partial  X}{\partial \sigma}\bigg) d\sigma}+\pi R_i^2 (U_\infty,V_\infty),
\end{equation}
in the $\sigma-\tau$ plane. Substituting (\ref{eq9})-(\ref{eq:inverseP}) and (\ref{eq3}) into (\ref{eq:general_forces}) yields the force that acts on droplet $i$ due to the uniform flow and as well as the flow field interaction with the neighboring droplet $j$

\begin{equation} \label{eq:TwoDropsForces}
\begin{split}
    (F_{X,i},F_{Y,i})=-4 \pi A^2  \sum _{n=1}^{\infty} { n e^{-2 n |\tau _{i}|}\coth (n(|\tau_i| +|\tau_j|))}\cdot (U_i-U_\infty,V_i-V_\infty)+\\
    +4 \pi A^2 \sum _{n=1}^{\infty} {n(\coth(n(|\tau_i| +|\tau_j|))-1)}\cdot(U_j-U_\infty,-V_j+V_\infty) +\pi R_i^2 (U_\infty,V_\infty).
\end{split}
\end{equation}
For the case of a droplet within a larger droplet we obtain
\begin{equation} \label{eq:interactionDropInDrop}
\begin{split}
 (F_{X,i},F_{Y,i})=  -4 \pi A^2\sum _{n=1}^{\infty} { n e^{-2 n \tau _{i}} \coth (n |\tau_i -\tau_j|)}\cdot (U_i,V_i) \\-4 \pi A^2  \sum _{n=1}^{\infty} { \frac{n e^{- n (\tau _{i}+ \tau _{j})}}{\sinh(n|\tau_i -\tau_j|)}}\cdot (U_j,V_j),
 \end{split}
\end{equation}
where $(i,j)=(1,2)$ represents the external droplet, and $(i,j)=(2,1)$ represents the inner droplet. The forces are normalized with respect to the following characteristic force, $f^*=12 \mu_i u^* l^{*2}/g$.
Eqs. (\ref{eq:TwoDropsForces}), (\ref{eq:interactionDropInDrop}) allow for exact (in the Hele-Shaw limit) solution of interaction forces for arbitrarily chosen  droplet velocities. Force balance is required in order to examine the dynamics of two closely spaced droplets driven by external uniform flow. Since  (\ref{eq:TwoDropsForces}), (\ref{eq:interactionDropInDrop}) are normalized by the viscosity of the liquid in the calculated region (see Fig. \ref{fig:BiPolarCoordinates}b,c). For simplicity, we hereafter use dimensional parameters where $\mu_{1}$, $\mu_2$ and $\mu_s$ are the viscosities of droplet $1$, droplet $2$ and the surrounding liquid, respectively.



Commonly, experimental works estimate the total friction of the droplet by measuring the ratio $\beta=u_d/u_\infty$ \citep[e.g.][]{liu2012waves,Beatus2012103,shen2014dynamics}, where $u_d$ is the speed of an isolated droplet. The coefficient $\beta$ may be related to the internal friction force  by $f_{in}=-\mu_i \pi r_{2}^2 u_d-c \pi r_2^2 u_{d2}$, where $c=2 \mu_l/\beta-\mu_d-\mu_l$. In order to compute the velocities of the droplets, the external force acting on the droplet should be balanced with the internal friction. Given the dominant effect of capillary forces in micro-configurations, surface tension significantly affects the friction force of the droplet.





Applying the aforementioned force balance yields the equations of motion for two closely spaced droplets
\begin{subequations}\label{eq:velocity}
\begin{equation} \label{eq9222}
\frac{u_i}{u_{\infty}}= \frac{(\mu_{l} g_i- \mu_l g_{12}+ \mu_l r_i^2)(\mu_{dj} r_j^2+c r_j^2+\mu_l g_j)+\mu_l g_{12}(\mu_l g_j-\mu_l g_{12}+ \mu_l r_j^2)}{\mu_l^2 g_{12}^2+(\mu_{d1} r_1^2+c r_1^2+\mu_l g_1)(\mu_{d2} r_2^2+c r_2^2+\mu_l g_2)},
\end{equation}

\begin{equation} \label{eq9sfds}
\frac{v_i}{v_{\infty}}= \frac{(\mu_{l} g_i+ \mu_l g_{12}+ \mu_l r_i^2)(\mu_{dj} r_j^2+c r_j^2+\mu_l g_j)-\mu_l g_{12}(\mu_l g_j+\mu_l g_{12}+ \mu_l r_j^2)}{\mu_l^2 g_{12}^2+(\mu_{d1} r_1^2+c r_1^2+\mu_l g_1)(\mu_{d2} r_2^2+c r_2^2+\mu_l g_2)}.
\end{equation}
\end{subequations}
For a configuration of a droplet within a droplet (see Fig. \ref{fig:ModelDescribtion}b), the equations of motion are
\begin{subequations}\label{eq:aa}
\begin{equation}
\frac{u_1}{u_\infty}=\frac{v_1}{v_\infty}= \frac{2 r_1^2 \mu_l\mu_{d1} (-c  r_2^2-\mu_{d2}  r_2^2+ \mu_{d1} g_2)} {\mu_{d1}^2 g_{12}^2+(\mu_l r_1^2+c r_1^2 -\mu_{d1} g_1)(-\mu_{d2}  r_2^2-c  r_2^2+ \mu_{d1}g_2 )}
\end{equation}
\begin{equation}
\frac{u_2}{u_\infty}=\frac{v_2}{v_\infty}= \frac{-2 r_1^2 \mu_l \mu_d g_{12}} {\mu_{d1}^2 g_{12}^2+(\mu_l r_1^2+c r_1^2 -\mu_{d1} g_1)(-\mu_{d2}  r_2^2-c  r_2^2+ \mu_{d1}g_2 )}
\end{equation}
\end{subequations}
where $(u_1,v_1)$ represents the external droplet, and $(u_2,v_2)$ represents the inner droplet, 
\begin{subequations}\label{eq:dimen2_6a}
\begin{equation} \label{eq9dfsdfd}
g_{1}= 4 a^2 \sum _{n=1}^{\infty} { n e^{-2 n \tau _{1}} \coth (n(\tau_1 -\tau_2))}
,\quad g_{2}= 4 a^2 \sum _{n=1}^{\infty} { n e^{-2 n |\tau _{2}|} \coth (n(\tau_1 -\tau_2))},
\end{equation}
\begin{equation} \label{eq9bbbb}
g_{12}=4 a^2 \sum _{n=1}^{\infty} { \frac{-n e^{-2 n(\tau _{1}+ |\tau _{2}|)}}{\sinh (n(\tau_1 -\tau_2))}},
\end{equation}
and
\begin{equation}
a= \frac{\sqrt{h^2-(r_1+r_2)^2)(h^2-(r_1-r_2)^2)}}{2h}.
\end{equation}
\end{subequations}
We note that (\ref{eq:velocity}-\ref{eq:dimen2_6a}) are the droplet velocities for a coordinate system where $X$ is parallel to the line connecting the centers of the droplets. Thus, the coordinate system rotates with the motion of the droplets and hence $(u_\infty,v_\infty)$ needs to be recalculated accordingly.

\begin{figure}
\centering
\includegraphics[width=13.cm]{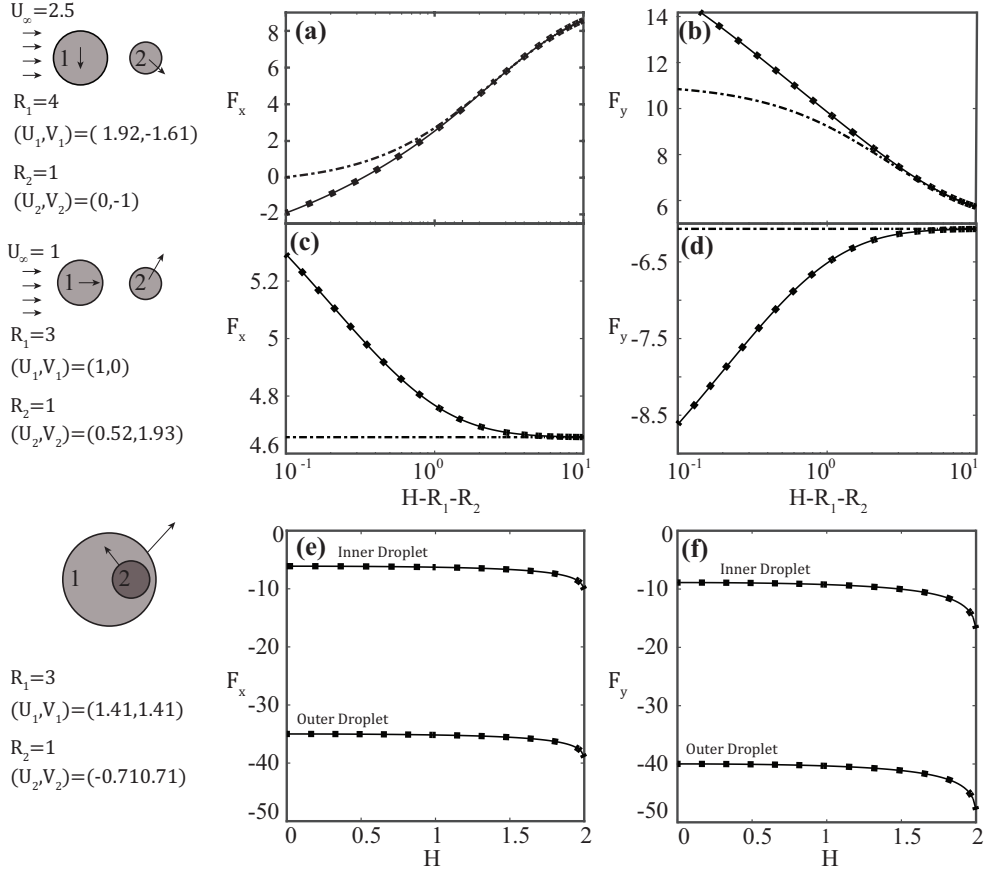}
\caption{Comparison between (\ref{eq:TwoDropsForces}-\ref{eq:interactionDropInDrop}) (smooth line), the dipole approximation (dashed line) and numerical calculations (dotted line). Panels $(a,c,e)$ present force in the $X$ direction vs. the distance between droplets. Panels $(b,d,f)$ present force in the $Y$ direction vs. the distance between droplets. Panels $(a-d)$ present the interaction forces of two adjacent droplets (\ref{eq:TwoDropsForces}). In panels (a,b)  $R_1 =1$,  $R_2 =4$, $(U_{1}, V_{1})=(1.92,-1.61)$, $(U_{2},V_{2})=(0,-1.5)$,  $(U_{\infty},V_{\infty})=(2.5,0)$. In panels (c,d) $R_1 =1$,  $R_2 =3$, $(U_{1},V_{1})=(0.52,1.93)$, $(U_{2},V_{2})=(1,0)$, and  $(U_{\infty},V_{\infty})=(1,0)$. Panels $(e,f)$ present the forces of a droplet within a droplet (\ref{eq:interactionDropInDrop}) where  $R_1 =3$,  $R_2 =1$, $(U_{1},V_{1})=(1.414,1.414)$, and $(U_{2},V_{2})=(-0.707,0.707)$.}
\label{fig:ForcesTwoCloseDroplets}
\end{figure}


Fig. $\ref{fig:ForcesTwoCloseDroplets}$ illustrates the forces acting on two closely spaced droplets given in (\ref{eq:TwoDropsForces}) (smooth lines, panels a-d), as well as the force acting on a droplet within a droplet configuration given in (\ref{eq:interactionDropInDrop}) (smooth lines, panels e-f). Both analytical solutions (\ref{eq:TwoDropsForces},\ref{eq:interactionDropInDrop}) are compared to the numerical solution of (\ref{laplace_tau_sigma}) (dotted lines). In the case of closely spaced droplets, the dipole approximation (dashed lines) is compared to the exact analytical solution (\ref{eq:TwoDropsForces}) as well. The forces in the $x$-direction (panels a,c,e) and the $y$-direction (panels b,d,f) are presented vs. the distance between the droplets $H-R_1-R_2$ in panels $(a-d)$ and vs. $H$ in panels $e,f$. In all cases, the analytical solutions given in (\ref{eq:TwoDropsForces}, \ref{eq:interactionDropInDrop}) are in the perfect agreement with the numerical solution of (\ref{laplace_tau_sigma})-(\ref{eq8}). In the case of $H-R_1-R_2\ll 1$, significant difference exists between the dipole model and the exact solution (\ref{eq:TwoDropsForces}). However this difference vanishes  for $H-R_1-R_2\gg 1$ as $H \rightarrow \infty$, and the force approaches the solution of a single isolated droplet, $\pi R_1^2(2  {U_\infty} -{U_1})$. In panels $(c,d)$ we illustrate the case of one droplet moving with the same velocity as the external flow and thus have dipole of zero strength. In this case $H-R_1-R_2$ does not affect the interaction forces. Panels (e,f) examine the forces acting on a droplet within a droplet, yielding approximately constant forces for $H<1$. Additionally, the interaction force increases with $H$ as the small droplet approaches the boundary of the bigger droplet.



\begin{figure}
\centering
\includegraphics[width=13cm ]{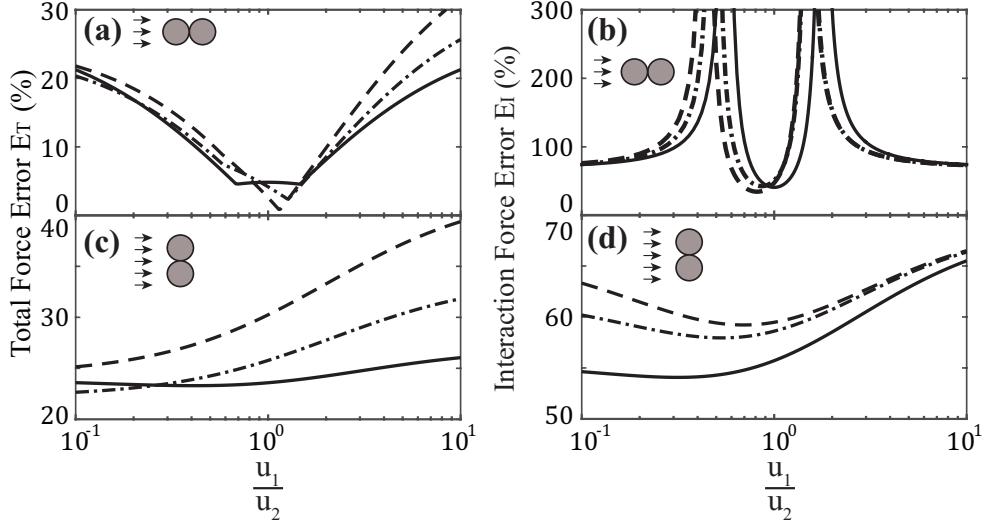}
\caption{Accuracy of the dipole approximation compared with the exact solution (\ref{eq:TwoDropsForces}). Panels $(a,c)$ present error of the dipole approximation scaled by the total force, $E_T=|F_{exact}-F_{dipole}|/|F_{d\infty}|$. Panels $(b,d)$ present the  error for the dipole approximation scaled by the interaction force, $E_R=|F_{exact}-F_{dipole}|/|F_{interaction}|$. In Panels $(a,b)$ the pair of droplets is parallel to the external flow. In Panels $(c,d)$ the pair of droplets is perpendicular to the external flow. $R_1/R_2=1$ (smooth line), $R_1/R_2=2$ (dashed dotted line), $R_1/R_2=4$ (dashed line), in all cases $U_\infty=5$.}
\label{fig:ErrorGraphs}
\end{figure}

Fig. \ref{fig:ErrorGraphs} illustrates the difference between (\ref{eq:TwoDropsForces}) and the dipole approximation for the limit of $H\rightarrow0$. In all cases $U_\infty=5$, and $R_1/R_2=1,2,4$ , corresponding to smooth, dashed and dashed-dotted lines, respectively. The line connecting the centers of the droplets is parallel (panels $a,b$) or perpendicular (panels $c,d$) to the uniform flow. Since (\ref{eq:TwoDropsForces}) is an exact solution (under the assumptions of the Hele-Shaw model), we define the  error of the dipole approximation by $E_T=|F_{exact}-F_{dipole}|/|F_{d\infty}|$ (panels $a,c$). For parallel configuration (panel $a$) clear minima are evident for $U_1/U_2\approx1$ where the error decreases to $\approx 5\%$. For the perpendicular configuration no minima is evident and the errors vary from $25\%$ to $40\%$. In fact, various theoretical and experimental models concerned with the dynamics of micro-fluidic droplets (e.g. excitation of the phonon modes in a droplet lattice \citep{beatus2006phonons} require the knowledge of the inter-droplet interaction forces. These interaction forces are represented by the difference between the total force and the force acting on an isolated droplet. In the present study we define the error of the interaction force as $E_R=|F_{exact}-F_{dipole}|/|F_{interaction}|$ presented in (panels $b,d$). The errors associated with the interaction forces are $\approx 100\%$ for the parallel configuration (panel b) and $\approx 60\%$ for the perpendicular configuration (panel d). As it can be inferred from the results of Fig. \ref{fig:ErrorGraphs}, the dipole approximation of the total force is more accurate for the parallel configuration (panel a) in comparison to the perpendicular one (panel c). However, in the case of the interaction forces the opposite is true, namely the dipole approximation yields the more accurate approximation for the perpendicular configuration (panel d) compared to the parallel one (panel b).


\begin{figure}  
\centering
\includegraphics[width=11.cm]{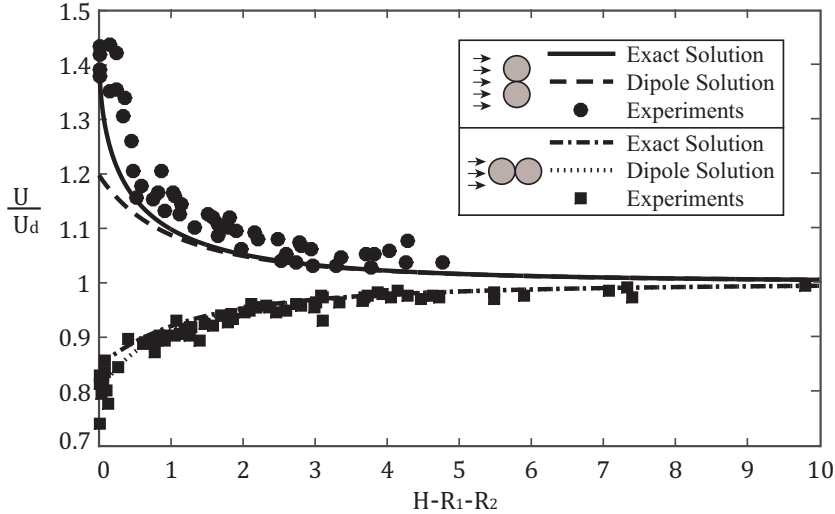}
\caption{Scaled speed of droplet pairs $U/U_d$ vs. the distance between the droplets $H-R_1-R_2$. $U_d$ is the speed of a single isolated droplet. The line connecting the centers of the droplets is: I. parallel to external flow (squares, dotted line and dotted-dashed line) or II. perpendicular to the external flow (circles, smooth line and dashed line). The experimental data was taken from \cite{shen2014dynamics} and are marked by squares and circles. Exact solution (\ref{eq:TwoDropsForces}) is marked by smooth and dashed-dotted lines. Dipole approximation is marked by dashed and dotted lines.}
\label{fig:experiments}
\end{figure}

In Fig. \ref{fig:experiments} we compare the experimental results (squares, circles) presented in \cite{shen2014dynamics} with the two analytical models, namely the classical dipole model (dotted line, dashed line)  and the exact model (\ref{eq:TwoDropsForces}) (smooth line, dash-dotted line) developed herein. In this figure the speed of droplet pairs are plotted vs. the distance between the droplets for the two distinct orientations: I. The line connecting the centers of the droplets is parallel to external flow (squares, dotted line and dotted-dashed line) and II. perpendicular to the external flow (circles, smooth line and dashed line). In case I. the configuration is close to the region of minimal difference between the dipole model and the exact solution  (\ref{eq:TwoDropsForces}) (see Fig. $\ref{fig:ErrorGraphs}$). In this case the difference between the models is negligible compared to the experimental resolution. However, in case II. (\ref{eq:TwoDropsForces}) accurately captures the change in slope of the experimental data as $H\rightarrow 0$, and thus significantly outperforms the original dipole model.

\begin{figure}
\centering
\includegraphics[width=12.cm]{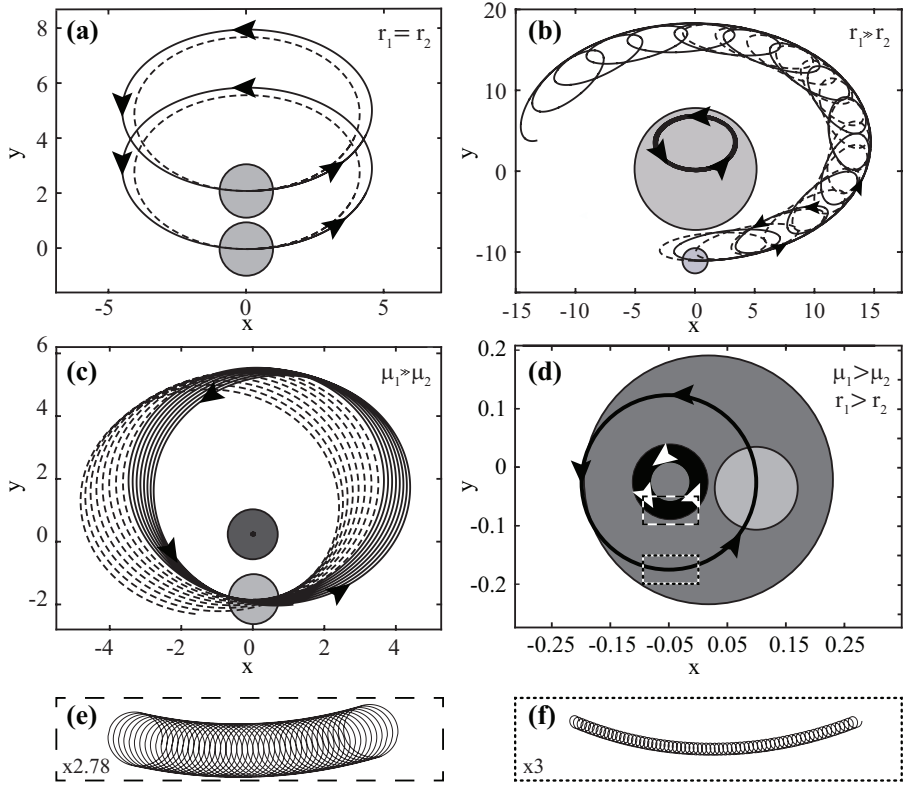}
\caption{Trajectories of the centers of two droplets due to an external rotating uniform velocity $(U_\infty,V_\infty)=|U_\infty|(\cos(\omega T),\sin(\omega T))$. Smooth lines are exact solution (\ref{eq:aa}-\ref{eq:dimen2_6a}) and dashed lines mark dipole approximation. Panels a-c, $|U_\infty|=0.4$, $\omega=0.025$. Panel a $R_1=R_2=1$, $\mu_1/\mu_2=1$. Panel b $R_1=10$, $R_2=1$ and $\mu_1/\mu_2=1$. Panel c $R_1=R_2=1$, $\mu_1/\mu_2=1000$. Panel d presents a droplet within a droplet where $|U_\infty|=0.5$, $\omega=1.256$, $R_1=3$, $R_2=1$ and $\mu_1/\mu_2=0.5$. Panels d, e present zoomed regions of the trajectories illustrated in Panel d.}

\label{fig:DynamicTwoDroplets}
\end{figure}

Fig. \ref{fig:DynamicTwoDroplets} presents the trajectories of the centers of two droplets subject to the uniform, rotating velocity field $(U_\infty,V_\infty)=|U_\infty|(\cos(\omega T),\sin(\omega T))$. Smooth lines denote calculations using the exact model (\ref{eq:aa},\ref{eq:dimen2_6a}) and dashed lines correspond to the original dipole model. Fig. \ref{fig:DynamicTwoDroplets}a illustrates the motion of two identical droplets ($R_1/R_2=1, \mu_1/\mu_2=1, |U_\infty|=0.4,  \omega=0.025$). By careful observation of the trajectories of both droplets shown in Fig. \ref{fig:DynamicTwoDroplets}a one can clearly see that due to the equal inter-droplet interaction forces the relative displacement between their centers remains permanently constant. However, the orbits of each droplet center are perfectly periodic and circulate around the fixed points in the plane.  It is worthwhile emphasising that the presence of the interaction forces brings about the modification of a single droplet trajectory. Fig. \ref{fig:DynamicTwoDroplets}b presents the trajectories of two droplets with different radii  ($R_1=1$ and $R_2=0.1$, $\mu_1/\mu_2=1, |U_\infty|=0.4, \omega=0.025$). In contrast to the previously considered case (i.e. identical droplets), the mismatch in the radii of two droplets introduces the asymmetry an the interaction forces applied on the droplets. In this case the relative distance between the centers of the droplets is time varying and two distinct (slow and fast) rotational components can be clearly identified. The fast $\omega$ component is manifested by the local, high frequency rotations drifting around the slowly evolving (slow component) outer orbit. In fact, the fast rotational component is induced by the external flow, whereas a slower frequency component can be attributed to the relative displacement between the centers of the droplets. Importantly, the original dipole model yields a significant error in approximation of the lower oscillatory mode. Trajectories shown in Fig. \ref{fig:DynamicTwoDroplets}c correspond to the case where the asymmetry is induced by the mismatch in viscosities $\mu_1/\mu_2=1000$ ($R_1/R_2=1$, keeping all other parameters identical). Similarly to the case shown in Fig. \ref{fig:DynamicTwoDroplets}b, the two distinct (slow and fast) rotational components are evident in Fig. \ref{fig:DynamicTwoDroplets}c. Interestingly enough, the dipole approximation yields the slow drift component in the opposite direction to that of the exact solution. Finally, in Fig. \ref{fig:DynamicTwoDroplets}d-f we considered a different, droplet within a droplet, configuration where the mismatch is introduced in both viscosities and radii  $R_1=3$ and $R_2=1$, again two distinct frequencies are evident. To clearly illustrate the low amplitude, fast scale rotations of both the terjactories  Fig. \ref{fig:DynamicTwoDroplets}d  we plot their local zooms in \ref{fig:DynamicTwoDroplets}e and Fig. \ref{fig:DynamicTwoDroplets}f. Zoomed regions are denoted with dashed rectangular frames in the inset of Fig. \ref{fig:DynamicTwoDroplets}d. 


\section{Concluding remarks}

We applied a transformation to a bi-polar coordinate system and obtained an exact solution for the interaction forces between two closely spaced droplets, as well as a droplet within a droplet, in a Hele-Shaw cell. The commonly used approximation by dipoles is shown to give significant errors ($10\%-30\%$) of the total force in the limit of small distance between the droplets. An exception is the case of two droplets moving at similar speeds, where the line connecting the centers of the droplets is parallel to the external uniform flow, yielding small error of $\approx 5\%$ or less. When we examine only the interaction forces, the errors are above $\approx 50\%$ and the dipole based model cannot be used to accurately capture the interaction forces.   

The analysis is performed in the framework of the Hele-Shaw limit, assuming negligible inertial effects and shallow configurations $g/l\rightarrow 0$. 
For cases in which the viscosity of the droplets is negligible compared with the viscosity of the computed region we expect errors of $(g/l)^2$ due to neglected viscous terms. For cases in which the viscosity of the droplets is similar or greater than the viscosity of the fluid in the computed region, the errors will increase to $g/l$ due to mismatch of the boundary conditions at the interface between the droplets and the surrounding fluid  \citep[see][]{gat2009higher}.



Let us close the current discussion with identifying several interesting future extensions of the current work. Due to the dominant effect of surface tension in micro-scale flows, small changes in the geometry of the droplet may induce significant capillary forces, while keeping the droplet geometry in the $X-Y$ plane approximately circular. Thus, the classic problem of multiple solutions for the motion of droplets and fingers in a Hele-Shaw cell \citep{kopf1988bubble} is still relevant to cases with dominant surface tension. Current experimental works resolve this by empirically estimating the speed of an isolated droplet. In addition, we believe that the solution presented in this work can be applied for the broad class of problems concerning the dynamics of the densely packed 1D and 2D droplet lattice structures. The newly proposed unit cell model of droplet within a droplet paves way for the future theoretical and experimental extensions to the weakly nonlinear phononic structures where the overall response of the lattice is highly affected by the internal dynamics of the inner elements. Nonlinear wave phenomena emerging in these quite novel acoustic structures is a subject of the current research and will be reported in the publications next to come.

\bibliographystyle{jfm}


\end{document}